# Kerker condition for enhancing emission rate and directivity of single emitter coupled to dielectric metasurfaces

Megha Khokhar[1], Faraz A Inam[2], and Rajesh V Nair[1*]

[1] *Laboratory for Nano-scale Optics and Meta-materials (LaNOM)*
*Department of Physics, Indian Institute of Technology Ropar, Punjab 140001 India*
[2]*Department of Physics, Aligarh Muslim University, Aligarh, Uttar Pradesh 202002, India*
*email: rvnair@iitrpr.ac.in*

## Abstract

Metasurfaces have the ability to control classical and non-classical states of light to achieve controlled emission even at the level of single emitter. Here, we unveil the Kerker condition induced emission rate enhancement with strong directivity from a single emitter integrated within a dielectric metasurface consists of silicon nano-disks. The simulation and analytical calculations attest the Kerker condition with unidirectional light scattering evolved by the constructive interference between electric dipole, toroidal dipole, and the magnetic quadrupole. The results evince spatially-dependent enhanced local density of optical states which reciprocates localized field intensity. The rate enhancement of 400 times is achieved at zero phonon line of nitrogen-vacancy center with superior emission directivity and collection efficiency. The results have implications in on-demand single photon generation, spin-photon interface, many-body interactions, and strong coupling.

Wave-scattering is a ubiquitous process not limited to light-matter interactions but applicable to condensed matter systems, nuclear physics, and atmospheric sciences [1,2]. The measure of wave scattering is quantified using the scattering cross-section, which dictate the interaction strength between the incoming particle and scatterers. Scattering optimization is a sought-after goal in atomic spectroscopy, nuclear scattering, and light-matter interactions. Specifically, the reliable control over light scattering results in photonic band gap, random lasing, wave-front shaping, and metamaterials that induce an exploration in fundamental science and applications [3]. The scattering optimization plays a pivotal role in modulating the emission dynamics of quantum emitters and spin-photon entanglement using pre-designed spatial-arrangement of sub-wavelength scatterers [4].

The generation of on-demand single photons and spins using solid-state emitters has refined scattering optimization using photonic metamaterials. The recent surge disseminates that negatively charged nitrogen-vacancy (NV-) center in diamond is a room-temperature single photon source with high quantum efficiency and spin coherence time [5]. The NV- center is formed by a substitutional nitrogen atom adjacent to a carbon vacancy in a diamond which finds applications in quantum sensing, magnetometry, and bio-markers [5,6]. The emission spectrum of NV- center consists of a pure electronic transition (zero phonon line (ZPL)) at 640 nm assisted with broad phonon sideband (PSB) emission [7]. Such PSB transitions induce decoherence with limited (1-2%) emission at ZPL even at low temperatures. However, it is necessary to enhance emission intensity and rate at the ZPL with simultaneous PSB inhibition for efficient use of NV- center in single photon generation, optical spin readout, and quantum sensing. Such reliable control over the emission requires the deterministic tuning of local density of optical states (LDOS). This demand scattering optimization using photonic metamaterials consisting of scatterers with different shapes, geometry, or their compositions [8-10]. This has realized unique phenomenon's in atomic antennas [11], metamaterials [12], and light emission [13]. The scattering optimization excites different types of electric and magnetic dipole moments that induce directional light scattering with suppression in either forward or backward scattering [14,15]. The complete suppression of backward scattering is known as the Kerker condition that occurs as a result of interference between the electric (*ED*) and magnetic dipole (*MD*) [16]. The role of higher-order modes is studied and that show overlapped electric (*EQ*) and magnetic quadrupole (*MQ*) modes [17]. The Kerker condition is generalized to different domains that include controlled atomic transitions [11], Huygens sources [18], spin-orbit interactions [19], and invisibility [20]. The dielectric metasurface with directional scattering can significantly modify the quantum emitter properties through LDOS tuning [21]. Hence we anticipate the ZPL enhancement with simultaneous PSB inhibition using Kerker condition; a long-awaited goal in NV center-based quantum technology [5].

In this Letter, we discuss emission rate enhancement of a single NV- center at ZPL wavelength of 640 nm exploiting the Kerker condition in dielectric metasurface. The constructive interference of several multipoles is evoked to obtain the Kerker condition, which is discussed using simulations supported by analytical calculations. The strong field confinement along with unidirectional scattering enable the tuning of emission rate. An enhancement of 400 times in the emission rate at ZPL wavelength is achieved with strong emission directivity and collection efficiency.

Figure 1(a) shows two-dimensional (2D) Silicon (Si) nano-disks arrays to achieve Kerker condition at 640 nm using finite-difference time-domain (FDTD) simulation (Lumerical). The optimized parameters of lattice constant (*a*) 310 nm, diameter (*d*) 230 nm, and thickness (*t*) 115 nm on a glass substrate of index 1.5 is used to obtain Kerker condition. The wavelength ($\lambda$)-dependent Si refractive index is used from the Palik handbook [22]: $n_0(\lambda) + i\beta(\lambda)$, where $n_0(\lambda)$ and $\beta(\lambda)$ are real and imaginary part of index, and thus losses associated with Si is included in the calculations. An *x*-polarized plane wave with wave vector ***k*** in *z*-direction induces electric and magnetic multipolar resonances decomposed using multipole expansion [23]. Figure 1(b) depicts schematic of dipolar resonance consists of electric and magnetic dipoles originated due to induced polarization. The oscillating electric dipole with dipole moment (***P***) create a current loop with density (***j***), which drives displacement current loop with magnetic dipole moment (***M***). The circulating current loop ***j*** in the axial direction constitute poloidal currents with toroidal dipole moment (***T***). These dipolar terms can interfere in constructive or destructive ways, at specific conditions, to achieve Kerker condition as shown in Fig. 1(b).

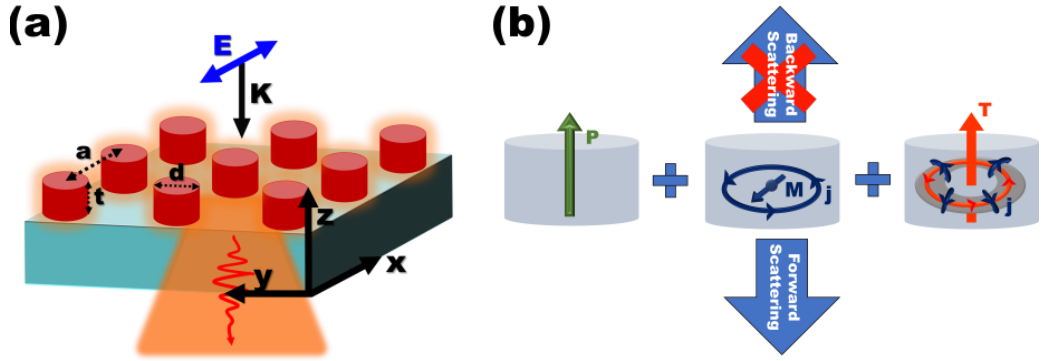

FIG. 1. (a) Schematic of metasurface consists of square array of Si nano-disks with $d$ = 230 nm, $t$ = 115 nm, and $a$ = 310 nm on a glass with incident $x$-polarized plane wave. (b) The constructive interference between the electric (***P***), magnetic (***M***), and toroidal (***T***) dipole moments results in forward scattering with null backward scattering.

Figure 2(a) shows simulated far-field transmission (reflection) spectra from an array of Si nano-disks that exhibit a peak (trough) at 640 nm. This peak originates due to interference between excited multipoles within nano-disk and their interaction with adjacent nano-disks [24,25]. The spectra show 12% reflectivity with 46% transmittance that illustrates a reduced backward scattering at 640 nm, as expected at Kerker condition. However, 42% of light is absorbed by the Si nano-disks that inhibit a complete forward scattering. A full transparency window with 100% light transmission at 640 nm is obtained for metasurfaces made with lossless material ($\beta$ = 0) with $\lambda$-dependent $n_0$ value, as seen in the inset of Fig. 2(a). Our results substantiate that null backward scattering is restricted by finite absorption (Supplementary material SI). The simulation results are corroborated with analytical calculations using multipole expansion method (Supplementary material SI).

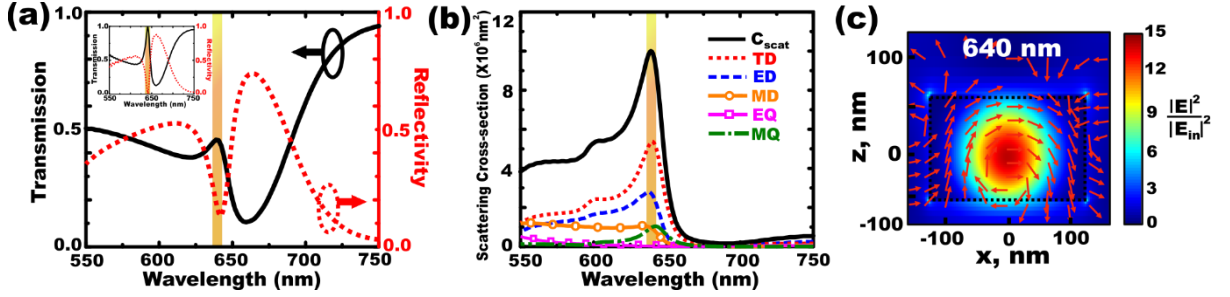

FIG. 2. (a) The calculated transmission (solid line) and reflectivity (dotted line) spectra with peak and trough demonstrates Kerker condition at 640 nm. The inset shows complete transparency at 640 nm with $\beta = 0$. (b) The contribution of each multipole in the scattering cross-section calculated using multipole expansion method. (c) The normalized field intensity distribution $\frac{|E|^2}{|E_{in}|^2}$ at 640 nm in *x-z* plane.

We estimate cartesian multipoles contributing to total scattering cross-section ($C_{scat}$) using FDTD and multipole expansion method [30]. A plane wave source excites the metasurface and electric field and refractive index profile is obtained in three-directions using FDTD simulations. These profiles are used in the modified MATLAB script to calculate the decomposed multipoles and $C_{scat}$. The induced displacement current density is: $\boldsymbol{j}(\boldsymbol{r}) = -i\omega\varepsilon_0(\varepsilon - \varepsilon_d)\boldsymbol{E}(\boldsymbol{r})$, where $\boldsymbol{E}(\boldsymbol{r})$ is field at position vector $\boldsymbol{r}$, $\omega = 2\pi f$ is angular frequency, $\varepsilon$ and $\varepsilon_0$ are the permittivity of metasurface and free space with $\varepsilon_d = 1$ as homogenous medium. The Kerker condition arises due to the constructive interference between the *ED* and *MD* having same phase and amplitude with dipole moments $\boldsymbol{P}$ and $\boldsymbol{M}$, given by: $\boldsymbol{P_\alpha} = \frac{-1}{i\omega} \int \boldsymbol{j}_\alpha d^3\boldsymbol{r}$ and $\boldsymbol{M_\alpha} = \frac{1}{2} \int (\boldsymbol{r} \times \boldsymbol{j})_\alpha d^3\boldsymbol{r}$ with $\alpha = (x,y,z)$ [23]. The poloidal current generates *TD* contribution in the scattering spectra given by $\boldsymbol{T_\alpha} = \frac{1}{10c} \int [(\boldsymbol{r} \cdot \boldsymbol{j})\boldsymbol{r} - 2r^2 \boldsymbol{j}_\alpha] d^3\boldsymbol{r}$. The higher order multipoles such as *EQ* and *MQ* also contributes to scattering, expressed as:

$$\boldsymbol{EQ}_{\alpha\beta} = \frac{-1}{i\omega} \int [3\,(r_\beta j_\alpha + r_\alpha j_\beta) - 2(\boldsymbol{r}\cdot\boldsymbol{j})\delta_{\alpha\beta}]d^3\boldsymbol{r} + \frac{k^2}{14}\int[4\,r_\alpha r_\beta(\boldsymbol{r}\cdot\boldsymbol{j}) - 5r^2(r_\alpha j_\beta + r_\beta j_\alpha) + 2r^2(\boldsymbol{r}\cdot\boldsymbol{j})\delta_{\alpha\beta}]d^3\boldsymbol{r} \quad (1)$$

$$\boldsymbol{MQ}_{\alpha\beta} = \int[r_\alpha(\boldsymbol{r}\times\boldsymbol{j})_\beta + r_\beta(\boldsymbol{r}\times\boldsymbol{j})_\alpha]d^3\boldsymbol{r} \quad\quad\quad \ldots\ldots (2)$$

where $\delta_{\alpha\beta}$ is Kronecker delta function with $\delta_{\alpha\beta} = 1$ for $\alpha = \beta$ and $\delta_{\alpha\beta} = 0$ otherwise, with $\alpha, \beta = x, y, z$, and $k = \omega/c$ where $c$ is speed of light in vacuum. Thus, scattered field is sum of contribution from all multipoles to $C_{scat}$ given as: $C_{scat} = ED + TD + MD + EQ + MQ$, which after substitution gives [23]:

$$C_{scat} = \frac{k^4}{6\pi\varepsilon_0^2 E_0}\left[|\boldsymbol{P}_\alpha|^2 + (ik)^2|\boldsymbol{T}_\alpha|^2 + \left|\frac{M_\alpha}{c^2}\right|^2 + \frac{1}{120}k^2|\boldsymbol{EQ}_{\alpha\beta}|^2 + \frac{1}{120}\frac{k^2}{c^2}|\boldsymbol{MQ}_{\alpha\beta}|^2\right] \quad\ldots\ldots (3)$$

The observed transmission peak at 640 nm in Fig. 2(a) corresponds to enhanced forward scattering arising due to multipole interference. Hence, it is necessary to calculate the multipolar contributions to the spectral-dependent $C_{scat}$ using Eq. (3) [26]. Figure 2(b) shows the superposition (solid line) of scattered fields arises from different types of excited multipoles. The calculated $C_{scat}$ is peaked at 640 nm, similar to the transmission peak. Conventionally, Kerker condition is explored as an overlap of electric and magnetic dipoles only, neglecting higher order terms [27]. However, we observed that *TD* (dotted line), *ED* (dashed line), and *MQ* (dash-dot-dash line) are the dominant multipoles contributing to $C_{scat}$ at 640 nm, as seen in Fig. 2(b). The contribution from *MD* (line with circles) and *EQ* (line with squares) is quite minimal. Hence, our results correspond to generalized Kerker effect originated due to the contributions from *TD*, *ED*, and *MQ* modes. The *ED* and *TD* scattered fields are analyzed using their phases and amplitudes (Supplementary material SII). The nearly same phase confirms the constructive interference between the *ED* and *TD* creating a super-dipole effect, which collectively constitutes a total electric dipole moment (*TED*). This is in contrast to anapole-like mode where destructive interference between *ED* and *TD* is observed [28]. Similarly, the *TED* and *MQ* modes constructively interfere with each other, while a destructive interference between the *TED* and *MD* modes is seen in the scattering spectra that inhibit forward scattering. Figure 2(c) shows the normalized electric field intensity distribution $\frac{|E|^2}{|E_{in}|^2}$ where $|E|^2$ is field intensity at 640 nm in comparison to the incident field intensity $|E_{in}|^2$. The cross-sectional (*x-z* plane) view shows an enhanced intensity by a factor of 15 at nano-disk center and then reduces to the edges. Such a field enhancement is obtained due to constructive interference of several multipoles with optimized scattering that can impart significant modification in the emission rate of single NV-center.

The spontaneous emission rate of a dipole emitter with transition dipole moment (**d**) depends on the environment through LDOS [29]. The strong optical field induced by excitation of electric and magnetic multipoles is related to LDOS ($\rho_d$) that measures the emission decay rate ($\Gamma$) at a position $\boldsymbol{r_0}$ as $\Gamma = \frac{\pi \omega_0 \mu_0}{\hbar} |d|^2 \rho_d(\boldsymbol{r_0}, \omega_0)$. Here $\rho_d(\boldsymbol{r_0}, \omega_0) = \frac{2\omega_0}{\pi c^2} [n. Im\{\boldsymbol{G}(\boldsymbol{r}, \boldsymbol{r_0}, \omega_0)\}. n]$ and $\boldsymbol{G}(\boldsymbol{r}, \boldsymbol{r_0}, \omega_0)$ is Dyadic Green's function which represent the electric field at $\boldsymbol{r}$ due to a dipole at $\boldsymbol{r_0}$ [30,31]. A significant field enhancement is achieved at the Kerker condition in Fig. 2(c) and hence, we anticipate an increased LDOS at 640 nm (corresponds to $\omega_0$), resulting in large decay rate enhancement. It is quite often reported that emission intensity enhancement is taken as a measure of dipoles excitation in metasurfaces [32]. However, emission intensity enhancement is a convoluted result of several parameters including the excitation pump intensity, variation in photonic environment, enhanced detection efficiency, and quenching effect. Hence, variation in intensity is an insufficient criterion to claim that LDOS is enhanced due to Kerker condition. Thus, emission rate calculation is necessary to ascertain emission enhancement due to Kerker condition.

Figure 3(a) shows the calculated wavelength-dependent relative LDOS (symbols) for a single emitter (single NV- center in a nanodiamond with index 2.45). This single emitter act as a point dipole source, driven by an external source, placed at each nano-disk center. Figure 3(a) inset shows spatial-dependent variation of $|E(\boldsymbol{r_0})|^2$ with an enhancement of 2000 at the nano-disk center. This intensity enhancement corroborates the enhanced field at nano-disk center which reduces to periphery as shown in Fig. 2(c). The LDOS for an emitter in metasurface with decay rate ($\Gamma$) is normalized to the same emitter placed in vacuum with a decay rate ($\Gamma_0$). The calculated wavelength-dependent relative decay rate (line) shows a significant increase at 640 nm, following LDOS enhancement at Kerker condition. This estimated rate enhancement ($\Gamma/\Gamma_0$) of 400 times with narrow emission bandwidth of 40 nm at room temperature is achieved due to spectral overlap between ZPL and Kerker condition at 640 nm. The increase in $\rho_d$ is due to constructive interference between the *ED*, *TD,* and *MQ* which enhances the rate of NV- ZPL using Kerker condition. Thus, the amplitude and phase of multipoles that result in constructive interference effectively modifies the local field intensity and impacts emission rate.

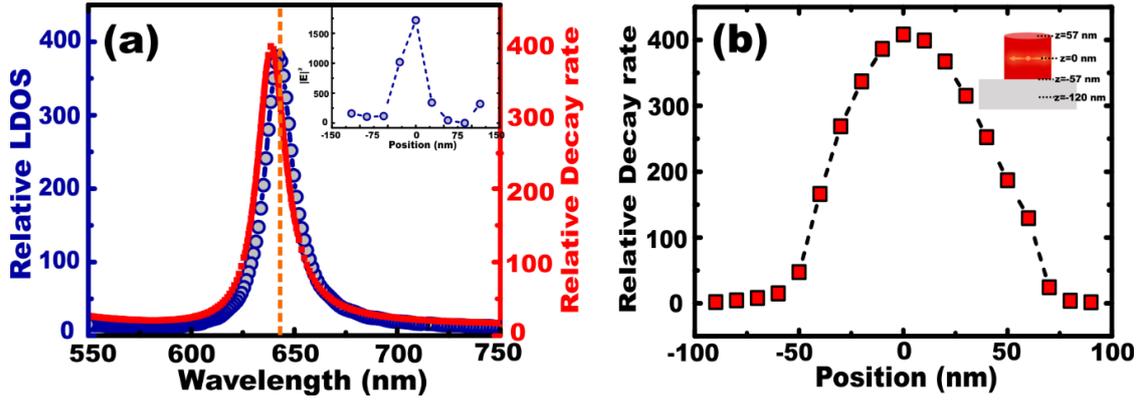

FIG. 3. (a) The calculated relative decay rate enhancement (line) for the emitter embedded inside the metasurface corresponding to LDOS maxima (symbols) at Kerker condition. The inset shows the spatial-dependent electric field intensity within the nano-disk. (b) The spatial-dependent rate enhancement for the dipole emitter emitting at 640 nm placed within the nano-disk (inset).

The excited multipolar moments constructively interfere with each other and further modifies the point dipole radiation. Therefore, the far-field radiation pattern of the quantum emitter can now be considered to be modified radiation pattern due to interaction with the induced multipolar moments of metasurface. Thus, the single dipole source can be treated as a single quantum emitter. The results confirm the ZPL rate enhancement with 100 nm bandwidth PSB suppression, which find applications in single photon emission with high brightness and useful for spin-photon interfaces. The increase in emission rate is similar to the enhancement due to an optical cavity, quantified using Purcell effect [33]. The Purcell enhancement is estimated to be 280 times for single NV- center integrated with metasurface (Supplementary material SIII). The results confirm strong enhancement in emission decay rate using Kerker condition which is a signature of LDOS enhancement.

Figure 3(b) depicts the calculated spatial-dependent decay rate enhancement at 640 nm with maximum enhancement is for an emitter placed at nano-disk center ($z$=0) as reciprocated by the spatial-dependent localized field. The calculated rate enhancement shows the mirror symmetry with respect to $z = 0.5t$. This enhancement shows an abrupt reduction when emitter position is shifted to $z = \pm 57$ *nm* (nano-disk edge), which follows the decrease in field intensity as seen in Fig. 2(c). The enhancement is reduced to 100 for emitter at nano-disk surface and reaches zero inside the substrate ($z < -57$ *nm*) or outside the metasurface ($z > 57$ *nm*). Thus, single emitter position within metasurface is crucial to achieve high emission rate at Kerker condition.

In addition to rate enhancement, emission directionality from single emitter manoeuvred by the excited multipoles is an important factor to be considered. Figure 4 shows calculated 2D angle-dependent far-field emission radiation pattern from the single emitter coupled to the metasurface in *x-z* plane. The 2D radiation pattern ($S(\theta, \varphi)$) is calculated using angular distribution of emitted power $P(\theta, \varphi, \lambda_e)$ as [34]:

$$S(\theta, \varphi) = \frac{P(\theta, \varphi, \lambda_e)}{Max(P_0(\theta, \varphi, \lambda_e))} \frac{|E|^2}{|E_{in}|^2} \quad \ldots\ldots\ldots\ldots(4)$$

where $P_0(\theta, \varphi, \lambda_e)$ is the emitted power in vacuum at emission wavelength $\lambda_e$, $\theta$, $\varphi$ are the polar and azimuthal angles, respectively. The directional nature of emission results from the constructive interference between emitter and excited multipoles. The plane-wave excitation is expected to result in an angular distribution mainly restricted to the forward scattering direction of the propagating plane wave at Kerker condition. The angular distribution observed in general is quite broad and extend over the whole forward hemisphere [35-37]. Figure 4(a) shows the angular distribution of the normalized radiation intensity for horizontal in-plane dipole, as the single emitter, placed on top of a glass substrate (red line) and in vacuum (green line) at 640 nm. The emitter does not consider forward or backward scattering and emits evenly in all-directions. However, emitted radiation for emitter on substrate is symmetrically confined in $z$-direction and antisymmetric along $x$-direction. This asymmetric radiation pattern is due to high-index glass substrate at bottom and air on top.

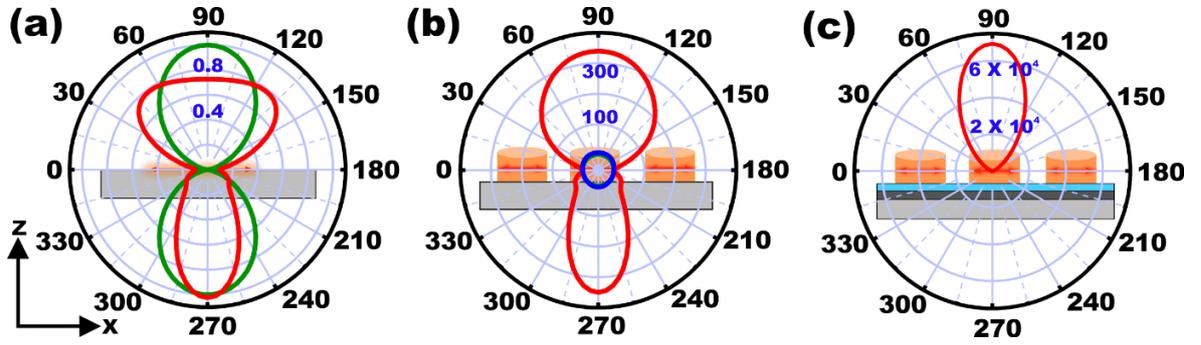

FIG. 4. The angular distribution of normalized radiation intensity in $x$-$z$ plane for (a) emitter on a glass substrate (red line) in contrast to vacuum (green line). (b) Emitter at the center of nano-disk radiating at 640 nm (red line), 540 nm (blue line), 740 nm (green line). (c) The enhanced radiation intensity at 640 nm for metasurface on top of reflector layer (grey colour).

Figure 4(b) shows the symmetric radiation pattern along $z$-direction for emitter placed at the nano-disk center. The radiation lobe in the lower half ($180° < \theta < 360°$) is restricted to smaller angular range in comparison to the upper half ($0° < \theta < 180°$). Since, the spontaneous emission from a radiating emitter is independent of excitation process, both vertical directions (top and bottom) are expected to be symmetric. However, the observed asymmetry in radiation pattern arises due to presence of the glass substrate and air. We notice an enhancement of ~300 in the far-field radiation intensity (normalized to maximum vacuum intensity) at 640 nm (red line). However, the enhancement is obtained only perpendicular to emitter axis within a limited angular range. This indicates that the emission pattern direction is induced by the constructive interference between emitted field from the emitter and scattered field by the *TED* and *MQ*. The radiation pattern is symmetric with *z*-axis due to even parity of *TED* and *MQ* modes [16]. The radiation pattern at 540 nm (blue line) and 740 nm (green line) shows an enhancement by a factor of 2 and 8 with an asymmetric emission radiation pattern due to glass substrate.

The directional emission can be further enhanced with a reduced angular range using a reflector layer beneath the metasurface. The silver layer with thickness 100 nm act as a reflector that enhances the radiation intensity to a value of $7\times10^4$ in the reflection geometry as shown in Fig 4(c). Further, the collection efficiency ($\eta$) is calculated as $I_{collection}/I_{total}$ with $I_{collection}$ is the radiated emission intensity in presence of metasurface and $I_{total}$ is radiated intensity over 360° solid angle. The calculated collection efficiency is 70% at 640 nm using a collection objective of numerical aperture 0.9 (solid angle: 64.15°) without the reflector layer from the metasurface. The efficient coupling to a single mode fiber could be substantially improved by considering metasurface on top of the reflector layer with very large emission directionality with near-unity collection efficiency of 99.9% using similar collection geometry. A comparison between our findings and those reported earlier in the broader areas of an optical cavity design is given in Supplementary Information SIV. The proposed metasurface at Kerker condition provides better collection efficiency, directivity, and Purcell enhancement for single NV- center.

Since, the quantum emitter's radiation pattern corresponds to a radiating point dipole, it is reported that the emission exhibits weak directionality and low collection efficiency [38]. Our studies demonstrate that quantum emitter integrated with metasurfaces provide high directionality while improving its collection efficiency using Kerker condition. The observed modification in emission rate is not limited to single NV centers in diamond, rather it can be generalized to variety of emitters in SiC and h-BN [39,40]. The proposed results are further applicable to achieve low-threshold NV- lasing. The Kerker mode switching can be achieved by a rapid change in refractive index through optical Kerr effect or free carrier generation [41]. Such Kerker based optical switching, combined with single NV-, enables deterministic single photon source. Further, with incorparation of excitonic materials, the strong coupling for excitons can be achieved using Kerker condition [42]. The Kerker condition is tunable through the appropriate choice of lattice constants and materials so that it can be achieved at any frequencies. This has remarkable use in the better readout of NV- spins with an enhanced sensitivity and high-fidelity using the localized micro-wave field at Kerker condition. Such Kerker based spin readout can be an alternative approach to the conventional micro-wave cavity based approach [43]. The 2D array of atoms is recently proposed as quantum metasurface and the excitation of multipole moments can lead to complete forward scattering in such atomic arrays [44]. Thus, our results encompass broad domains of interest and applicable to any kind of wave-wave scattering in an appropriate medium.

To conclude, we have studied Kerker condition induced enhancement in the emission rate of single NV- center at the ZPL wavelength of 640 nm while suppressing the non-resonant PSB emission. A complete forward scattering is obtained at the Kerker condition using simulations supported by analytical calculations. The Kerker condition is originated due to the constructive interference between *ED* and *TD*, which forms a super-dipole effect combined with the contribution from *MQ*. Further, decay rate enhancement of 400 times is obtained at Kerker condition with a strong directivity and collection efficiency. The rate enhancement depends on emitter position within nano-disk vis-a-vis the localized field intensity with maximum rate enhancement at nano-disk center. Our results are substantial to reinforce the research in quantum metasurfaces by exploiting the Kerker condition for generating on-demand single photon source, quantum imaging, many-body interactions, and strong coupling between emitters.

**Acknowledgements**

The authors would like to acknowledge the financial support from DST-RFBR [INT/RUS/RFBR/P-318], DST-ICPS [DST/ICPS/QuST/Theme-2/2019/General], DST-SERB [SB/SJF/2020-21/05]. RVN acknowledges the Swarnajayanti Fellowship (DST/SJF/PSA-01/2019-20).

[1] D. Belkic, *Principles of quantum scattering theory* (CRC Press, 2020).
[2] C. F. Bohren and D. R. Huffman, *Absorption and scattering of light by small particles* (John Wiley & Sons, 2008).
[3] M. Ghulinyan and L. Pavesi, *Light Localisation and Lasing: Random and Pseudo-random Photonic Structures* (Cambridge University Press, 2015).
[4] A. S. Solntsev, G. S. Agarwal, and Y. S. Kivshar, Metasurfaces for quantum photonics, Nature Photonics **15**, 327 (2021).
[5] M. W. Doherty, N. B. Manson, P. Delaney, F. Jelezko, J. Wrachtrup, and L. C. Hollenberg, The nitrogen-vacancy colour centre in diamond, Phys. Rep. **528**, 1 (2013).
[6] L. Rondin, J.-P. Tetienne, T. Hingant, J.-F. Roch, P. Maletinsky, and V. Jacques, Magnetometry with nitrogen-vacancy defects in diamond, Rep. Prog. Phys. **77**, 056503 (2014).
[7] S. Sharma and R. V. Nair, Nanophotonic control of the color center emission from nanodiamonds, Opt. Lett. **43**, 3989 (2018).
[8] S. Bidault, M. Mivelle, and N. Bonod, Dielectric nanoantennas to manipulate solid-state light emission, J. Appl. Phys. **126** (2019).
[9] H. Choi, D. Zhu, Y. Yoon, and D. Englund, Cascaded cavities boost the indistinguishability of imperfect quantum emitters, Phys. Rev. Lett. **122**, 183602 (2019).
[10] H. Kaupp *et al.*, Purcell-enhanced single-photon emission from nitrogen-vacancy centers coupled to a tunable microcavity, Phys. Rev. Appl. **6**, 054010 (2016).
[11] R. Alaee, A. Safari, V. Sandoghdar, and R. W. Boyd, Kerker effect, superscattering, and scattering dark states in atomic antennas, Phy. Rev. Res. **2**, 043409 (2020).
[12] S. Kruk and Y. Kivshar, Functional Meta-Optics and Nanophotonics Governed by Mie Resonances, ACS Photonics **4**, 2638 (2017).
[13] A. Vaskin, R. Kolkowski, A. F. Koenderink, and I. Staude, Light-emitting metasurfaces, Nanophotonics **8**, 1151 (2019).
[14] L. Cong and R. Singh, Spatiotemporal dielectric metasurfaces for unidirectional propagation and reconfigurable steering of terahertz beams, Adv. Mater. **32**, 2001418 (2020).
[15] A. B. Evlyukhin, S. M. Novikov, U. Zywietz, R. L. Eriksen, C. Reinhardt, S. I. Bozhevolnyi, and B. N. Chichkov, Demonstration of magnetic dipole resonances of dielectric nanospheres in the visible region, Nano Lett. **12**, 3749 (2012).
[16] W. Liu and Y. S. Kivshar, Generalized Kerker effects in nanophotonics and meta-optics, Opt. express **26**, 13085 (2018).
[17] A. Pors, S. K. Andersen, and S. I. Bozhevolnyi, Unidirectional scattering by nanoparticles near substrates: generalized Kerker conditions, Opt. express **23**, 28808 (2015).
[18] M. Decker, I. Staude, M. Falkner, J. Dominguez, D. N. Neshev, I. Brener, T. Pertsch, and Y. S. Kivshar, High-Efficiency Dielectric Huygens' Surfaces, Adv. Opt. Mater. **3**, 813 (2015).
[19] J. Olmos-Trigo, C. Sanz-Fernández, D. R. Abujetas, A. García-Etxarri, G. Molina-Terriza, J. A. Sánchez-Gil, F. S. Bergeret, and J. J. Sáenz, Role of the absorption on the spin-orbit interactions of light with Si nano-particles, J. Appl. Phys. **126**, 033104 (2019).
[20] X. Zhang and A. L. Bradley, Wide-angle invisible dielectric metasurface driven by transverse Kerker scattering, Phys. Rev. B **103**, 195419 (2021).
[21] R. Alaee, R. Filter, D. Lehr, F. Lederer, and C. Rockstuhl, A generalized Kerker condition for highly directive nanoantennas, Opt. Lett. **40**, 2645 (2015).
[22] E. D. Palik, *Handbook of optical constants of solids* (Academic press, 1998), Vol. 3.
[23] R. Alaee, C. Rockstuhl, and I. Fernandez-Corbaton, An electromagnetic multipole expansion beyond the long-wavelength approximation, Opt. Commun. **407**, 17 (2018).
[24] V. E. Babicheva and J. V. Moloney, Lattice effect influence on the electric and magnetic dipole resonance overlap in a disk array, Nanophotonics **7**, 1663 (2018).

[25] H. K. Shamkhi, A. Sayanskiy, A. C. Valero, A. S. Kupriianov, P. Kapitanova, Y. S. Kivshar, A. S. Shalin, and V. R. Tuz, Transparency and perfect absorption of all-dielectric resonant metasurfaces governed by the transverse Kerker effect, Phys. Rev. B **3** (2019).

[26] T. Hinamoto and M. Fujii, MENP: an open-source MATLAB implementation of multipole expansion for nanophotonics, OSA Continuum **4**, 1640 (2021).

[27] I. Staude, A E. Miroshnichenko, M. Decker, N. T. Fofang, S. Liu, E. Gonzales, J. Dominguez, Luk TS Luk, DN Neshev, I Brener, and Y Kivshar, Tailoring directional scattering through magnetic and electric resonances in subwavelength silicon nanodisks, ACS nano **7**, 7824 (2013).

[28] J. A. Parker, H. Sugimoto, B. Coe, D. Eggena, M. Fujii, N. F. Scherer, S. K. Gray, and U. Manna, Excitation of Nonradiating Anapoles in Dielectric Nanospheres, Phys. Rev. Lett. **124**, 097402 (2020).

[29] M. Khokhar and R. V. Nair, Observation of finite-size-induced emission decay rates in self-assembled photonic crystals, Phys. Rev. A **102**, 013502 (2020).

[30] P. Lodahl, A. F. Van Driel, I. S. Nikolaev, A. Irman, K. Overgaag, D. Vanmaekelbergh, and W. L. Vos, Controlling the dynamics of spontaneous emission from quantum dots by photonic crystals, Nature **430**, 654 (2004).

[31] W. L. Vos, A. F. Koenderink, and I. S. Nikolaev, Orientation-dependent spontaneous emission rates of a two-level quantum emitter in any nanophotonic environment, Phys. Rev. A **80**, 053802 (2009).

[32] I. Staude, V. V. Khardikov, N. T. Fofang, S. Liu, M. Decker, D. N. Neshev, T. S. Luk, I. Brener, and Y. S. Kivshar, Shaping Photoluminescence Spectra with Magnetoelectric Resonances in All-Dielectric Nanoparticles, ACS Photonics **2**, 172 (2015).

[33] A. Krasnok, S. Glybovski, M. Petrov, S. Makarov, R. Savelev, P. Belov, C. Simovski, and Y. Kivshar, Demonstration of the enhanced Purcell factor in all-dielectric structures, Appl. Phys. Lett. **108** (2016).

[34] Z. Xi, Y. Lu, W. Yu, P. Yao, P. Wang, and H. Ming, Tailoring the directivity of both excitation and emission of dipole simultaneously with two-colored plasmonic antenna, Opt. express **21**, 29365 (2013).

[35] R. Alaee, R. Filter, D. Lehr, F. Lederer, and C. Rockstuhl, A generalized Kerker condition for highly directive nanoantennas, Opt. lett. **40**, 2645 (2015).

[36] S. Person, M. Jain, Z. Lapin, J. J. Sáenz, G. Wicks, and L. Novotny, Demonstration of zero optical backscattering from single nanoparticles, Nano Lett. **13**, 1806 (2013).

[37] H. K. Shamkhi *et al.*, Transverse scattering and generalized kerker effects in all-dielectric mie-resonant metaoptics, Phys. Rev. Lett. **122**, 193905 (2019).

[38] D. Le Sage, L. M. Pham, N. Bar-Gill, C. Belthangady, M. D. Lukin, A. Yacoby, and R. L. Walsworth, Efficient photon detection from color centers in a diamond optical waveguide, Phys. Rev. B **85**, 121202 (2012).

[39] S. Castelletto, Silicon carbide single-photon sources: challenges and prospects, Materials for Quantum Technology **1**, 023001 (2021).

[40] T. T. Tran, K. Bray, M. J. Ford, M. Toth, and I. Aharonovich, Quantum emission from hexagonal boron nitride monolayers, Nat. Nanotechnol. **11**, 37 (2016).

[41] S. V. Makarov, A. S. Zalogina, M. Tajik, D. A. Zuev, M. V. Rybin, A. A. Kuchmizhak, S. Juodkazis, and Y. Kivshar, Light-induced tuning and reconfiguration of nanophotonic structures, Laser & Photonics Rev. **11**, 1700108 (2017).

[42] G. W. Castellanos, S. Murai, T. Raziman, S. Wang, M. Ramezani, A. G. Curto, and J. Gómez Rivas, Exciton-polaritons with magnetic and electric character in all-dielectric metasurfaces, ACS Photonics **7**, 1226 (2020).

[43] E. R. Eisenach, J. F. Barry, M. F. O'Keeffe, J. M. Schloss, M. H. Steinecker, D. R. Englund, and D. A. Braje, Cavity-enhanced microwave readout of a solid-state spin sensor, Nat. commun. **12**, 1 (2021).

[44] R. Bekenstein, I. Pikovski, H. Pichler, E. Shahmoon, S. F. Yelin, and M. D. Lukin, Quantum metasurfaces with atom arrays, Nat. Phys **16**, 676 (2020).

Supplementary Material

# Kerker condition for enhancing emission rate and directivity of single emitter coupled to dielectric metasurfaces

Megha Khokhar[1], Faraz A Inam[2], and Rajesh V Nair[1*]

[1] *Laboratory for Nano-scale Optics and Meta-materials (LaNOM)
Department of Physics, Indian Institute of Technology Ropar, Punjab 140001 India*
[2] *Aligarh Muslim University, Aligarh, Uttar Pradesh, India*

\* Email: rvnair@iitrpr.ac.in

## SI. Absorption effect and analytical calculations of Kerker condition

Figure 2(a) in the main manuscript shows the Kerker condition for an array of silicon (Si) nano-disks with diameter 230 nm, thickness 115 nm, and periodicity 310 nm. The wavelength ($\lambda$) - dependent refractive index of Si is taken as: $n_0(\lambda) + i\beta(\lambda)$ where $n_0(\lambda)$ and $\beta(\lambda)$ are the wavelength-dependent real and imaginary part of index. A constant $\beta$ value is used in the spectral range of interest as its variation is minimal while a wavelength-dependent $n_0$ is included in the calculation, as seen in Fig. S1(a,b).

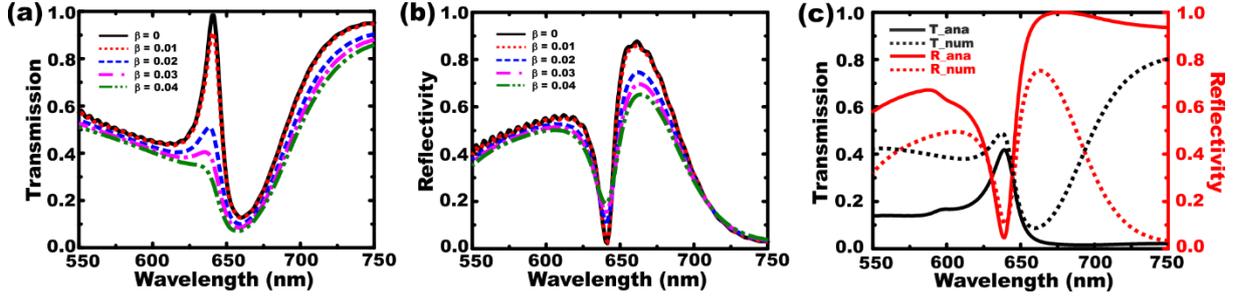

Figure S1. (a) Transmission and (b) reflectivity spectra for metasurface consisting of Si nano-disks on a glass substrate. The plot shows variation in transmission peak (reflectivity dip) values corresponding to the Kerker condition with imaginary part of the refractive index ($\beta$). (c) The transmission (black lines) and reflectivity (red lines) spectra for the metasurface calculated through analytical (solid lines) and numerical simulations (dotted lines).

The peak transmission value is unity with nearly zero reflectivity value at 640 nm for $\beta = 0$ (solid line), which shows a complete suppression of backward scattering is possible using lossless material. The transmission peak value steadily decreases with an increase in the $\beta$ value and similarly, the decrease in amplitude is seen for reflectivity dip. This implies that the induced losses (absorption) in the material impose the constraints in achieving the perfect forward scattering while suppressing complete backward scattering. At $\beta = 0.03$, the transmission peak at 640 nm is difficult to differentiate from the background spectra and it diminishes with further increase in the value of $\beta$.

In principle, to demonstrate the obtained results behind our study, we could have chosen any ambiguous high refractive index dielectric with zero absorption. However, those results could never be realised or practically achieved in fabricated samples. Si material is purposely chosen for this study due to its relatively high refractive index and minimal absorption (*n* ~ 3.87 and *β* ~ 0.018) at the ZPL wavelength of NV- center at 640 nm. The Si scatterers owns high scattering efficiency due to its high refractive index. Hence, Si is widely used for studying higher order electric/magnetic Mie-scattering moments of dielectric resonators comprising metasurface [1]. Also, due to its large-scale use in micro, nanofabrication for a wide range of electronic/photonic applications, the matured fabrication technology and precision for Si structure growth is far-advanced relative to other high refractive index materials. Therefore, for application purpose, metasurfaces based on Si can easily be fabricated. As an alternative to Si, other semiconductors like gallium arsenide (GaAs) or gallium nitride (GaN) could also be used due to their high refractive indices. However, the absorption (*β*) values for GaAs are similar to that of Si in the interested wavelength range. Alternatively, materials like diamond and silicon carbide (SiC) which are nearly perfect dielectrics with an almost zero *β* value, have relatively low-refractive index 2.41 and 2.63 at 640 nm, respectively. For such moderate refractive index materials, the multipolar resonance exhibit peak only due to electric dipole, as well as higher order multipolar moments disappears due to the increased overlap between the various Mie scattering moments [2]. Also, the quadrupolar or higher order moments are substantially reduced [2]. Further, fabricating the diamond or SiC based metasurfaces is very complicated process and hence, Si-based metasurfaces hold supremacy in the wavelength region of interest.

The analytical calculations are performed with an incident plane wave source, polarized in *x*-direction that excites the multipoles in the nano-disks with an effective electric $\boldsymbol{p}_E$ and magnetic polarizabilities $\boldsymbol{p}_M$. The component of the electric and magnetic dipole moment is given as $\boldsymbol{P}_x = \varepsilon_0 \varepsilon_d p_E E_0$ and $\boldsymbol{M}_y = p_M H_0$ where $E_0$ and $H_0$ are the electric and magnetic components, $\varepsilon_0$ and $\varepsilon_d$ is the permittivity of free space and surrounding medium, respectively. The forward and backward scattered fields are obtained from multipole moments (mentioned in main manuscript) and the transmission (*t*) and reflection (*r*) coefficients are estimated as [3]:

$$t = 1 + \frac{ik_d}{2E_0 A \varepsilon_0 \varepsilon_d} \left( \boldsymbol{P}_x + \frac{\boldsymbol{M}_y}{c} + \boldsymbol{T}_x - \frac{ik_d}{6} \boldsymbol{EQ}_{xz} - \frac{ik_d}{2c} \boldsymbol{MQ}_{yz} \right) \quad \text{(S1)}$$

$$r = \frac{ik_d}{2E_0 A \varepsilon_0 \varepsilon_d} \left( \boldsymbol{P}_x - \frac{\boldsymbol{M}_y}{c} + \boldsymbol{T}_x + \frac{ik_d}{6} \boldsymbol{EQ}_{xz} - \frac{ik_d}{2c} \boldsymbol{MQ}_{yz} \right) \quad \text{(S2)}$$

where *A* is the simulation region which is equal to square of the lattice period, and $k_d$ is the surrounding medium wave vector with $\varepsilon_d = 1$ and thus, $k_d = k_0$, wave vector in vacuum. The transmission and reflectivity are further calculated as $T = |t|^2$ and $R = |r|^2$, respectively. Figure S1(c) shows the analytically calculated transmission (T_ana) and reflectivity (R_ana) spectra. The analytically calculated spectra (solid line) are in complete agreement with the corresponding simulated (dotted lines) spectra with a sharp transmission peak (reflectivity dip) at Kerker condition. However, the spectral features differ at the longer wavelengths due to the variation in the extracted scattered fields contributed by the multipoles, observed in *xyz*-directions for the analytical calculations while *x-y* plane in simulations.

**SII. Analysis of multipole phases**

Figure S2(a) shows nearly the same phase induced by the electric dipole (*ED*) and toroidal dipole (*TD*) with dipole moments $P_x$ and $T_x$, respectively which suggest the constructive interference between these dipolar moments at 640 nm. This results in total electric dipole (*TED*) moment with $P_x - ikT_x = 0$ [4]. Since, the suppression of backward scattering with complete forward scattering occurs when $P_x$ and magnetic dipoles ($M_y$) are shifted by $\pi$ phase, for an incident *x*-polarized plane wave. The electric quadrupole (*EQ$_{xz}$*) also has a phase difference of $\pi$ with respect to the magnetic quadrupole (*MQ$_{yz}$*). Thus, the phases can be estimated as [4]:

$$\arg(P_x) = \arg(M_y) + 2\pi n; \; n \text{ is an integer} \quad (S3)$$

$$\arg(EQ_{xz}) = \arg(MQ_{yz}) + 2\pi n; \; n \text{ is an integer} \quad (S4)$$

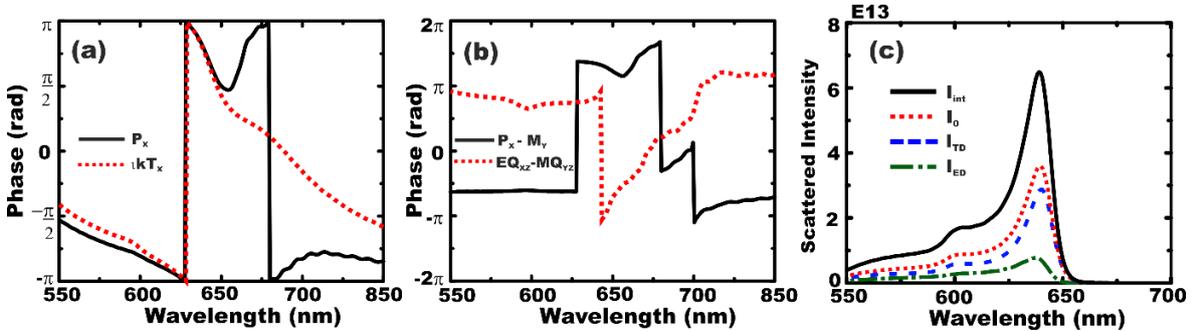

Figure S2. (a) Phase of the multipole contributions with electric (*ED*) and toroidal dipole (*TD*) moments shown with solid and dotted lines, respectively. (b) Phase relation for the dipoles (solid line) and quadrupoles (dotted lines). (c) The comparison of the scattering intensities for *TD* and *ED* component as $I_{TD}$ and $I_{ED}$ and their contribution as a sum of their intensities $I_0$ and superposition of their intensities as $I_{int}$.

Figure S2(b) shows the phase relations for dipoles (solid line) with a phase difference of nearly $2\pi$ at 640 nm. This satisfies the "dipole Kerker condition" as seen in Eq. (S3). We have also obtained a $2\pi$ phase difference for the quadrupoles (dotted line) which shows a sudden drop in the phase from $\pi$ to $-\pi$ at the Kerker condition, which validates Eq. (S4) [4]. We have observed constructive interference between the dipoles and quadrupoles that enhances the forward scattering with formalism: $\arg(\boldsymbol{P}_x - \boldsymbol{M}_y) = \arg(E\boldsymbol{Q}_{xz} - M\boldsymbol{Q}_{yz}) + 2\pi n$ (S5)

Thus, at the Kerker condition, the phase of dipole is π while the quadrupoles are out of phase by -π, which enhance the forward scattering.

Further, scattered intensity is calculated to unveil the contribution of multipoles at the Kerker condition. Since the forward scattering is mainly dominated by the *ED* and *TD*, the corresponding scattering intensity for *ED* ($I_{ED}$) and *TD* ($I_{TD}$) is calculated with $I_0$ as the sum of their intensity while $I_{int}$ is the superposition of both *ED* and *TD* modes [5]:

$$I_0 = I_{ED} + I_{TD} \quad ; \quad |I_{int}| = |I_{TD}|^2 + |I_{ED}|^2 + 2|I_{TD}||I_{ED}|\cos(\phi_{TD} - \phi_{ED}) \quad \text{(S6)}$$

where $\phi_{TD}$ and $\phi_{ED}$ are the phase of the *TD* and *ED* components. Fig. S2(c) shows the calculated intensities of *TD* and *ED* multipoles with $I_{TD} > I_{ED}$, indicating *TD* has main contribution in the Kerker condition. Also, $I_{int} > I_0$, which implies constructive interference between the *TD* and *ED* contributing to the *TED*. We have also observed a constructive interference between the *TED* and *MQ*, in addition to destructive interference of *MD*. Thus, the calculated scattering cross-section (Fig. 2(b) in main manuscript) results from the constructive interference between the *TD*, *ED*, and *MQ* at Kerker condition.

## SIII. Purcell factor enhancement calculations

The spontaneous emission rate for dipole emitter in the metasurface ($\Gamma$) relative to that in a free space ($\Gamma_0$) quantitatively estimates the Purcell factor. The interaction strength determining the Purcell enhancement is calculated as [6]: $F = \frac{\Gamma}{\Gamma_0} = F_P \frac{|\boldsymbol{u}(\boldsymbol{r}) \cdot \boldsymbol{d}|^2}{|\boldsymbol{d}|^2} \frac{1}{1 + \left(\frac{2Q}{\hbar\omega_c}\right)^2 (\hbar\omega_0 - \hbar\omega_c)^2}$

where $\boldsymbol{u}(\boldsymbol{r})$ is the electric field amplitude of the cavity mode at the position $\boldsymbol{r}$ and $\boldsymbol{d}$ is the transition dipole moment. The term $1/1 + \left(\frac{2Q}{\hbar\omega_c}\right)^2 (\hbar\omega_0 - \hbar\omega_c)^2$ describes the spectral dependency of the Purcell factor. $F_P$ is the maximum achievable Purcell factor that depends on the quality factor $Q$, mode volume of the cavity, $\omega_0$ as the emitter oscillation frequency, $\omega_c$ as the mode frequency and at resonance, $\omega_0 = \omega_c$. We have observed a Purcell enhancement (solid line) at $\omega_0$ corresponding to the resonant wavelength of 640 nm at the Kerker condition as shown in Fig. S3(a). The interference of multipoles enhances the localized field that further enhances the emission rate at the zero phonon line of NV- center. Figure S3(b) shows the maximum Purcell enhancement for the emitter located at the nano-disk center ($z = 0$) with an enhancement of 280 times as $\boldsymbol{u}(\boldsymbol{r}) = 1$ and $F = F_P$ at the Kerker condition. As the emitter position is varied from $z = 0$ on either side, the Purcell factor value decreases due to the reduction in localized field. Hence, positioning the single emitter within each nano-disk is crucial for obtaining the maximum Purcell enhancement.

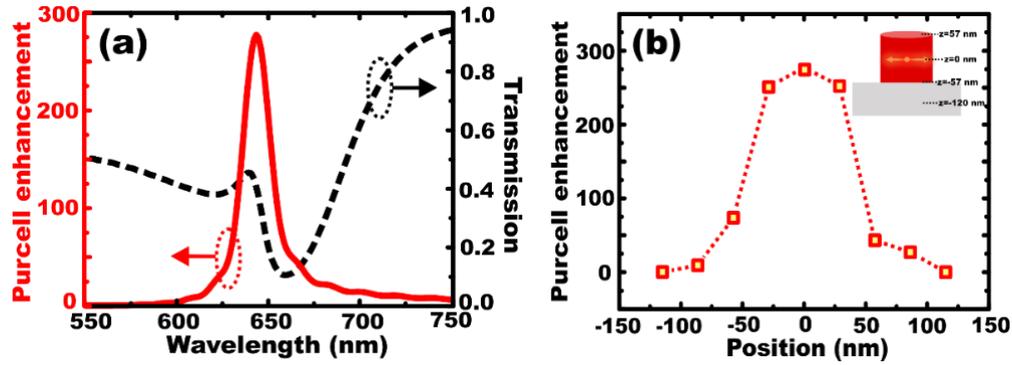

Figure S3. (a) The Purcell enhancement obtained for the emitter located at nano-disk center shows large emission corresponding to transmission peak at Kerker condition. (b) The fluctuation in enhancement factor with emitter located at different nano-disk positions.

## SIV.  Comparison table for different cavities

In this table, the figure of merits is compiled for the previous reports on the cavity type structures integrated with an emitter and are compared with present results.

| Cavity Type [Ref.] | Silicon-air Photonic bridge slot cavity [7] | Cascaded two-cavity system [8] | Circular Bragg Grating [9] | H1 photonic crystal-based GaAs membrane [10] | Silicon Nanowires (SiNW) [11] | Si nano-disks metasurface [Present manuscript] |
|---|---|---|---|---|---|---|
| **Active or Passive system** | Passive | Active (emitters: SiV centers) | Active (emitters: InAs/GaAs QDs) | Active (emitters: InAs QDs) | Active (emitters: MoS2 film) | Active (emitters: NV-center) |
| **Q Factor** | $> 10^6$ | First Cavity, $Q_1 = 7000$ Second Cavity, $Q_2 = 500000$ | .................... | 15000 | .................... | 80 |
| **Mode Volume** | $7.01 \times 10^{-5} \lambda^3$ | $0.007 (\lambda/n)^3$ | .................. | .................. | .................... | $2.17 \times 10^{-2} (\lambda/n)^3$ |
| **Emission Rate modification/Purcell enhancement** | Not discussed | Not discussed | Low Purcell enhancement =3 | No discussion on rate enhancement. | Absence of emission rate modification. | Purcell enhancement =280; Decay rate enhancement=400. Phonon sideband suppression |
| **Collection efficiency ($\eta$)** | .................... | With Indistinguishability as 0.95; ($\eta$) as 0.76% | 48% | 80% | Top/Bottom emission enhancement = 25. | $\eta > 99\%$ with back reflector $\eta > 70\%$ without back reflector |

| Additional Remarks | Fabricating 1nm bridge-slot cavity and single-emitter coupling therein is a challenge | Making cascaded cavities and mode alignment would be difficult. | Discussion on positioning of the emitters. | Experimentally feasible design | Emission intensity is used as a parameter to discuss enhancement. | Feasible design using the current state-of-the-art nano-fabrication. Emitter considered has both photonic and spin properties. Indistinguishability is not discussed. Kerker condition induced field localization. |

.............. Not mentioned/discussed in the paper

[1]     I. Staude and J. Schilling, Metamaterial-inspired silicon nanophotonics, Nat Photonics **11**, 274 (2017).
[2]     A. García-Etxarri, R. Gómez-Medina, L. S. Froufe-Pérez, C. López, L. Chantada, F. Scheffold, J. Aizpurua, M. Nieto-Vesperinas, and J. J. Sáenz, Strong magnetic response of submicron silicon particles in the infrared, Opt. Express **19**, 4815 (2011).
[3]     P. D. Terekhov, V. E. Babicheva, K. V. Baryshnikova, A. S. Shalin, A. Karabchevsky, and A. B. Evlyukhin, Multipole analysis of dielectric metasurfaces composed of nonspherical nanoparticles and lattice invisibility effect, Phys. Rev. B **99** (2019).
[4]     H. K. Shamkhi *et al.*, Transverse Scattering and Generalized Kerker Effects in All-Dielectric Mie-Resonant Metaoptics, Phys. Rev. Lett. **122**, 193905 (2019).
[5]     X. Zhang, J. Li, J. F. Donegan, and A. L. Bradley, Constructive and destructive interference of Kerker-type scattering in an ultrathin silicon Huygens metasurface, Phys. Rev. Materials **4** (2020).
[6]     D. Englund, D. Fattal, E. Waks, G. Solomon, B. Zhang, T. Nakaoka, Y. Arakawa, Y. Yamamoto, and J. Vuckovic, Controlling the spontaneous emission rate of single quantum dots in a two-dimensional photonic crystal, Phys. Rev. Lett. **95**, 013904 (2005).
[7]     H. Choi, M. Heuck, and D. Englund, Self-similar nanocavity design with ultrasmall mode volume for single-photon nonlinearities, Phys. Rev. Lett. **118**, 223605 (2017).
[8]     H. Choi, D. Zhu, Y. Yoon, and D. Englund, Cascaded cavities boost the indistinguishability of imperfect quantum emitters, Phys. Rev. Lett. **122**, 183602 (2019).
[9]     L. Sapienza, M. Davanço, A. Badolato, and K. Srinivasan, Nanoscale optical positioning of single quantum dots for bright and pure single-photon emission, Nat. Commun. **6**, 1 (2015).
[10]    J. Hagemeier *et al.*, H1 photonic crystal cavities for hybrid quantum information protocols, Opt. Express **20**, 24714 (2012).
[11]    A. F. Cihan, A. G. Curto, S. Raza, P. G. Kik, and M. L. Brongersma, Silicon Mie resonators for highly directional light emission from monolayer MoS2, Nat. Photonics **12**, 284 (2018).

-----------------------------------------------------------------